\documentclass[preprint2]{aastex63}
\usepackage{float, bm, graphicx, amsmath, morefloats, enumitem}
\submitjournal{ApJ}

\shorttitle{AT\,2020blt}
\shortauthors{Ho et al.}

\definecolor{DarkOrange}{RGB}{204, 85, 0}
\definecolor{LincolnGreen}{RGB}{17, 102, 0}
\def\ion#1#2{#1$\;${\footnotesize\rm{#2}}\relax}

\newcommand{\swift}{{\em Swift}}
\newcommand{\fermi}{{\em Fermi}}

\newcommand{\erg}{\mbox{$\rm erg$}}
\newcommand{\kev}{\mbox{$\rm keV$}}

\newcommand{\km}{\mbox{$\rm km$}}

\newcommand{\pcmsq}{\mbox{$\rm cm^{-2}$}}

\newcommand{\degsq}{\mbox{$\rm deg^{2}$}}

\newcommand{\pcmcub}{\mbox{$\rm cm^{-3}$}}

\newcommand{\msol}{\mbox{$\rm M_\odot$}}

\newcommand{\days}{\mbox{$\rm d$}}
\newcommand{\psec}{\mbox{$\rm s^{-1}$}}
\newcommand{\pyr}{\mbox{$\rm yr^{-1}$}}

\newcommand{\ghz}{\mbox{$\rm GHz$}}
\newcommand{\hz}{\mbox{$\rm Hz$}}
\newcommand{\phz}{\mbox{$\rm Hz^{-1}$}}

\begin{document}

\title{ZTF20aajnksq (AT\,2020blt): A Fast Optical Transient at $z \approx 2.9$ \\ With No Detected Gamma-Ray Burst Counterpart}

\author[0000-0002-9017-3567]{Anna Y. Q.~Ho}
\affiliation{Cahill Center for Astrophysics, 
California Institute of Technology, MC 249-17, 
1200 E California Boulevard, Pasadena, CA, 91125, USA}
\affiliation{Department of Astronomy, University of California, Berkeley, 501 Campbell Hall, Berkeley, CA, 94720, USA}
\affiliation{Miller Institute for Basic Research in Science, 468 Donner Lab, Berkeley, CA 94720, USA}

\author[0000-0001-8472-1996]{Daniel A.~Perley}
\affiliation{Astrophysics Research Institute, Liverpool John Moores University, IC2, Liverpool Science Park, 146 Brownlow Hill, Liverpool L3 5RF, UK}

\author[0000-0001-7833-1043]{Paz Beniamini}
\affiliation{Cahill Center for Astrophysics,
California Institute of Technology, MC 249-17,
1200 E California Boulevard, Pasadena, CA, 91125, USA}

\author[0000-0003-1673-970X]{S.\ Bradley~Cenko}
\affiliation{Astrophysics Science Division, NASA Goddard Space Flight Center, Mail Code 661, Greenbelt, MD 20771, USA}
\affiliation{Joint Space-Science Institute, University of Maryland, College Park, MD 20742, USA}

\author[0000-0001-5390-8563]{S. R.~Kulkarni}
\affiliation{Cahill Center for Astrophysics, 
California Institute of Technology, MC 249-17,
1200 E California Boulevard, Pasadena, CA, 91125, USA}

\author[0000-0002-8977-1498]{Igor Andreoni}
\affiliation{Cahill Center for Astrophysics, 
California Institute of Technology, MC 249-17, 
1200 E California Boulevard, Pasadena, CA, 91125, USA}

\author[0000-0003-1673-970X]{Leo P.~Singer}
\affiliation{Astrophysics Science Division, NASA Goddard Space Flight Center, Mail Code 661, Greenbelt, MD 20771, USA}

\author[0000-0002-8989-0542]{Kishalay De}
\affiliation{Cahill Center for Astrophysics, 
California Institute of Technology, MC 249-17,
1200 E California Boulevard, Pasadena, CA, 91125, USA}

\author{Mansi M.~Kasliwal}
\affiliation{Cahill Center for Astrophysics,
California Institute of Technology, MC 249-17,
1200 E California Boulevard, Pasadena, CA, 91125, USA}

\author[0000-0002-4223-103X]{Christoffer Fremling}
\affiliation{Cahill Center for Astrophysics, 
California Institute of Technology, MC 249-17, 
1200 E California Boulevard, Pasadena, CA, 91125, USA}

\author[0000-0001-8018-5348]{Eric C.~Bellm}
\affiliation{DIRAC Institute, Department of Astronomy, University of Washington, 3910 15th Avenue NE, Seattle, WA 98195, USA}

\author[0000-0002-5884-7867]{Richard Dekany}
\affiliation{Caltech Optical Observatories, California Institute of Technology, Pasadena, CA  91125}

\author{Alexandre Delacroix}
\affiliation{Caltech Optical Observatories, California Institute of Technology, Pasadena, CA  91125}

\author[0000-0001-5060-8733]{Dmitry A.~Duev}
\affiliation{Cahill Center for Astrophysics, 
California Institute of Technology, MC 249-17, 
1200 E California Boulevard, Pasadena, CA, 91125, USA}

\author[0000-0003-3461-8661]{Daniel A.~Goldstein}
\altaffiliation{Hubble Fellow}
\affiliation{Cahill Center for Astrophysics, 
California Institute of Technology, MC 249-17, 
1200 E California Boulevard, Pasadena, CA, 91125, USA}

\author[0000-0001-8205-2506]{V. Zach Golkhou}
\affiliation{DIRAC Institute, Department of Astronomy, University of Washington, 3910 15th Avenue NE, Seattle, WA 98195, USA} 
\affiliation{The eScience Institute, University of Washington, Seattle, WA 98195, USA}

\author[0000-0002-4163-4996]{Ariel Goobar}
\affiliation{The Oskar Klein Centre \& Department of Physics, Stockholm University, AlbaNova, SE-106 91 Stockholm, Sweden}

\author{Matthew J.~Graham}
\affiliation{Cahill Center for Astrophysics, 
California Institute of Technology, MC 249-17, 
1200 E California Boulevard, Pasadena, CA, 91125, USA}

\author{David Hale}
\affiliation{Caltech Optical Observatories, California Institute of Technology, Pasadena, CA  91125}

\author[0000-0002-6540-1484]{Thomas Kupfer}
\affiliation{Kavli Institute for Theoretical Physics, University of California, Santa Barbara, CA 93106, USA}

\author[0000-0003-2451-5482]{Russ R.~Laher}
\affiliation{IPAC, California Institute of Technology, 1200 E. California Blvd, Pasadena, CA 91125, USA}
             
\author[0000-0002-8532-9395]{Frank J.~Masci}
\affiliation{IPAC, California Institute of Technology, 1200 E. California Blvd, Pasadena, CA 91125, USA}

\author[0000-0001-9515-478X]{Adam~A.~Miller}
\affiliation{Center for Interdisciplinary Exploration and Research in Astrophysics (CIERA) and Department of Physics and Astronomy, Northwestern University, 1800 Sherman Road, Evanston, IL 60201, USA}
\affiliation{The Adler Planetarium, Chicago, IL 60605, USA}

\author[0000-0002-0466-1119]{James D.~Neill}
\affiliation{Cahill Center for Astrophysics, 
California Institute of Technology, MC 249-17, 
1200 E California Boulevard, Pasadena, CA, 91125, USA}

\author{Reed Riddle}
\affiliation{Caltech Optical Observatories, California Institute of Technology, Pasadena, CA  91125}

\author[0000-0001-7648-4142]{Ben Rusholme}
\affiliation{IPAC, California Institute of Technology, 1200 E. California Blvd, Pasadena, CA 91125, USA}

\author[0000-0003-4401-0430]{David L.~Shupe}
\affiliation{IPAC, California Institute of Technology, 1200 E. California
             Blvd, Pasadena, CA 91125, USA}

\author[0000-0001-7062-9726]{Roger Smith}
\affiliation{Caltech Optical Observatories, California Institute of Technology, Pasadena, CA  91125}

\author[0000-0003-1546-6615]{Jesper Sollerman}
\affiliation{The Oskar Klein Centre \& Department of Astronomy, Stockholm University, AlbaNova, SE-106 91 Stockholm, Sweden}

\author[0000-0002-2626-2872]{Jan van Roestel}
\affiliation{Cahill Center for Astrophysics, 
California Institute of Technology, MC 249-17, 
1200 E California Boulevard, Pasadena, CA, 91125, USA}

\begin{abstract}

We present ZTF20aajnksq (AT\,2020blt), a fast-fading ($\Delta r=2.3\,$mag in $\Delta t=1.3\,$d) red ($g-r\approx0.6\,$mag) and luminous ($M_{1626\,\text{\AA}}=-25.9\,$mag) optical transient at $z=2.9$
discovered by the Zwicky Transient Facility (ZTF).
AT\,2020blt shares several features in common with afterglows to long-duration gamma-ray bursts (GRBs):
(1) an optical light curve well-described by a broken power-law with a break at $t_\mathrm{j}=1\,$d (observer-frame);
(2) a luminous ($L_X = 10^{46}\,\mathrm{erg}\,\mathrm{s}^{-1}$) X-ray counterpart;
and (3) luminous ($L_\nu = 4 \times 10^{31}\,\erg\,\psec\,\phz$ at 10\,\ghz) radio emission.
However, no GRB was detected in the 0.74\,d between the last ZTF non-detection ($r>21.36\,$mag) and the first ZTF detection ($r=19.60\,$mag),
with an upper limit on the isotropic-equivalent gamma-ray energy release of $E_{\gamma,\mathrm{iso}} < 7 \times 10^{52}\,$erg.
AT\,2020blt is thus the third afterglow-like transient discovered without a detected GRB counterpart (after PTF11agg and ZTF19abvizsw) and the second (after ZTF19abvizsw) with a redshift measurement.
We conclude that the properties of AT\,2020blt are consistent with a classical (initial Lorentz factor $\Gamma_0 \gtrsim 100$) on-axis GRB that was missed by high-energy satellites.
Furthermore, by estimating the rate of transients with light curves similar to that of AT\,2020blt in ZTF high-cadence data, we agree with previous results that there is no evidence for an afterglow-like phenomenon that is significantly more common than classical GRBs,
such as dirty fireballs.
We conclude by discussing the status and future of fast-transient searches in wide-field high-cadence optical surveys.

\end{abstract}

\section{Introduction}
\label{sec:introduction}

Over the past half-century,
thousands of long-duration gamma-ray bursts (GRBs; \citealt{Piran2004,Zhang2004,Meszaros2006,Kouveliotou2012}) have been discovered by high-energy satellites.
In the traditional GRB model, a collapsing massive star launches a collimated (opening angle $\theta_0 \approx 10^{\circ}$) and ultrarelativistic (initial Lorentz factor $\Gamma_0 \gg 100$) outflow \citep{MacFadyen1999} that tunnels through the stellar material and collides with the ambient medium, producing an ``afterglow'' across the electromagnetic spectrum \citep{vanParadijs2000,Panaitescu2002}.

Through follow-up observations of well-localized GRB triggers, hundreds of optical afterglows have been detected\footnote{An up-to-date list is maintained at \url{http://www.mpe.mpg.de/\~jcg/grbgen.html}}.
There are several reasons why optical surveys should also detect ``orphan'' afterglows,
i.e. optical afterglows without associated GRBs.
First, for an outflow with Lorentz factor $\Gamma$,
relativistic beaming precludes the observer from seeing emission outside a cone of width
$\theta = 1/\Gamma$.
The outflow decelerates between the time of the GRB detection and the time of the optical afterglow detection,
so the optical afterglow should be visible over a wider observing angle than the GRB \citep{Rhoads1997,Meszaros1998}.
Second, an outflow must entrain very little mass ($M_\mathrm{ej}\approx 10^{-5}\,\msol$) to produce a GRB.
If GRBs represent the extreme of a continuum of baryon-loading in relativistic jets,
then ``dirty fireballs'' should exist, which would produce an afterglow but not a GRB \citep{Dermer1999}.

To discover orphan afterglows and dirty fireballs, surveys must be able to find afterglows without relying on a GRB trigger.
Independently discovering optical afterglow emission is challenging because of the need for high-cadence observations over a wide field-of-view,
as well as rapid follow-up.
Furthermore, there is
a formidable fog of more common fast-fading transients like stellar flares \citep{Kulkarni2006,Rau2008,Berger2013,Ho2018,vanRoestel2019}.
Of the three optically discovered afterglows in the literature,
two turned out to have associated classical GRBs:
iPTF14yb \citep{Cenko2015} was the counterpart to GRB\,140226A,
and ATLAS17aeu \citep{Bhalerao2017,Stalder2017} was likely the counterpart to GRB\,170105A.\footnote{The association is not fully secure, because the redshift of the afterglow was not measured.}

The first optically discovered afterglow, PTF11agg \citep{Cenko2013}, had no detected GRB counterpart.
The redshift was constrained to be $1 < z < 2$,
and \citet{Cenko2013} argued that it could represent the first dirty fireball.
It has since become clear that the rate of such events is not significantly higher than the rate of classical GRBs \citep{Cenko2015,Ho2018};
the same conclusion was reached by \citet{Nakar2003} based on X-ray afterglows.
So, if dirty fireballs exist, they are either rare
or look significantly different from classical GRB afterglows.

Making the discovery of optical afterglows \emph{routine} is one of the primary scientific goals of the Zwicky Transient Facility (ZTF; \citealt{Graham2019,Bellm2019}) high-cadence surveys \citep{Bellm2019surveys}.
To that end, we have devised a set of alert-stream filters for identifying afterglow emission in real-time, and obtaining prompt follow-up observations to measure the redshift and any accompanying X-ray and radio emission.
Here we describe the first afterglow detected as part of this effort,
ZTF20aajnksq (AT\,2020blt) at $z \approx 2.9$.
Since then, we discovered ZTF20abbiixp (AT\,2020kym; \citealt{Ho2020gcn}), which turned out to be the afterglow to Fermi/LAT GRB\,200524A (Yao et al. in prep).
In September 2019, ZTF also serendipitously discovered a cosmological afterglow (ZTF19abvizsw at $z=1.26$; \citealt{Burdge2019,Ho2019gcn})
in follow-up observations of gravitional-wave trigger S190901ap (Perley et al. in prep).
Finally, ZTF detected the afterglow to GRB\,190106A as ZTF19aabgebm; the detection was in low-cadence data and therefore the transient did not pass the fast-transient filter.

This paper is organized as follows.
In \S \ref{sec:disc-basic-analysis} we present the discovery and follow-up observations of AT\,2020blt.
In \S \ref{sec:analysis} we model the outflow using the light curve and the spectral energy distribution (SED).
We discuss possible interpretations in \S \ref{sec:interpretation}, and conclude that we cannot rule out the possibility that AT\,2020blt was a classical GRB missed by high-energy detectors.
We summarize and look to the future in \S \ref{sec:conclusions}.

\section{Observations}
\label{sec:disc-basic-analysis}

\subsection{ZTF Discovery}
\label{sec:disc-ztfdiscovery}

The ZTF Uniform Depth Survey (Goldstein et al. in prep) covers $2000\,\degsq$ twice per night in $g$-, $r$-, and $i$-band using the 48-inch Samuel Oschin Schmidt telescope at Palomar Observatory (P48).
The ZTF observing system is described in \citet{Dekany2020}.
The pipeline for ZTF photometry makes use of the image subtraction algorithm of \citet{Zackay2016} and is described in \citet{Masci2019}.

AT\,2020blt was discovered at $r=19.57\pm0.14\,$mag (all magnitudes given in AB) in an image obtained on 2020 Jan 28.28\footnote{All times in this paper are given in UTC.},
at the position
$\alpha = 12^{\mathrm{h}}47^{\mathrm{m}}04.87^{\mathrm{s}}$, 
$\delta = +45^{\mathrm{d}}12^{\mathrm{m}}02.3^{\mathrm{s}}$ (J2000).
One and a half hours later,
the source had faded to $r=20.01\pm0.16\,$mag.

AT\,2020blt passed a filter that searches the ZTF alert stream \citep{Patterson2019} for young and fast transients. More specifically,
the filter identifies transients that:

\begin{itemize}
    \item have an upper limit from the previous night that is at least one magnitude fainter than the first detection,
    \item have no historical detections in the Catalina Real-Time Transient Survey \citep{Drake2009,Mahabal2011,Djorgovski2011}, ZTF, or the predecessor to ZTF the Palomar Transient Factory \citep{Law2009},
    \item have a real-bogus score $\texttt{drb}>0.9$, which is associated with a false positive rate of 0.4\% and a false negative rate of 5\% \citep{Duev2019},
    \item have a Galactic latitude $|b|>15$,
    \item have two detections separated by at least half an hour (to remove asteroids), and
    \item have no stellar counterpart ($\texttt{sgscore} < 0.76$; \citealt{Tachibana2018}).
\end{itemize}

AT\,2020blt fulfilled the criteria listed above:
the last upper limit from the high-cadence survey was 0.74\,d prior to the first detection with an upper limit of $r > 20.73\,$mag.
There is no source within 15\arcsec\ of the position of AT\,2020blt in ZTF $r$-band and $g$-band reference images,
with a 5-$\sigma$ limiting magnitude in the PSF-fit reference image catalog of $r=23.17\,$mag and $g=22.77\,$mag.

Motivated by the fast rise and lack of a detected host galaxy counterpart in ZTF reference images\footnote{The median magnitude of the TOUGH sample of 69 cosmological \emph{Swift} GRB host galaxies \citep{Hjorth2012} was $R=25.52\pm0.23\,$mag.}
we immediately triggered a series of follow-up observations (\S\ref{sec:followup}) which were coordinated
through the GROWTH ``Marshal'' \citep{Kasliwal2019}.
All observations will be made available on WISeREP,
the Weizmann Interactive Supernova Data Repository \citep{Yaron2012}.

\subsection{Follow-up Observations}
\label{sec:followup}

\subsubsection{Optical Imaging}
\label{sec:followup-imaging}

In searches for extragalactic fast transients,
the primary false positives are stellar flares in the Milky Way \citep{Kulkarni2006,Rau2008,Berger2013,Ho2018}.
At optical frequencies,
stellar flares can be distinguished from afterglow emission by color.
At peak, stellar flares have typical blackbody temperatures of $\sim 10,000\,$K \citep{Kowalski2013}, so optical filters will be on the Rayleigh-Jeans tail and
colors will obey $f_\nu \propto \nu^{+2}$ ($g-r=-0.17\,$mag).
By contrast, in optical bands synchrotron emission
will obey $g-r>0$.
For example, a spectral index of $f_\nu \propto \nu^{-0.7}$ \citep{Sari1998} corresponds to $g-r=0.24\,$mag.

To measure the color of AT\,2020blt,
we triggered target-of-opportunity (ToO) programs on the IO:O imager of the Liverpool Telescope\footnote{PI: D. Perley} (LT; \citealt{Steele2004})
and with the Spectral Energy Distribution Machine\footnote{PI: A. Ho} (SEDM; \citealt{Blagorodnova2018,Rigault2019}) on the automated 60-inch telescope at Palomar Observatory (P60; \citealt{Cenko2006}).
LT image reduction was provided by the basic IO:O pipeline. P60 and LT image subtraction
was performed following \citet{Fremling2016},
using PS1 images for $griz$ and SDSS for $u$-band.

\begin{figure*}
    \centering
    \includegraphics[width=\textwidth]{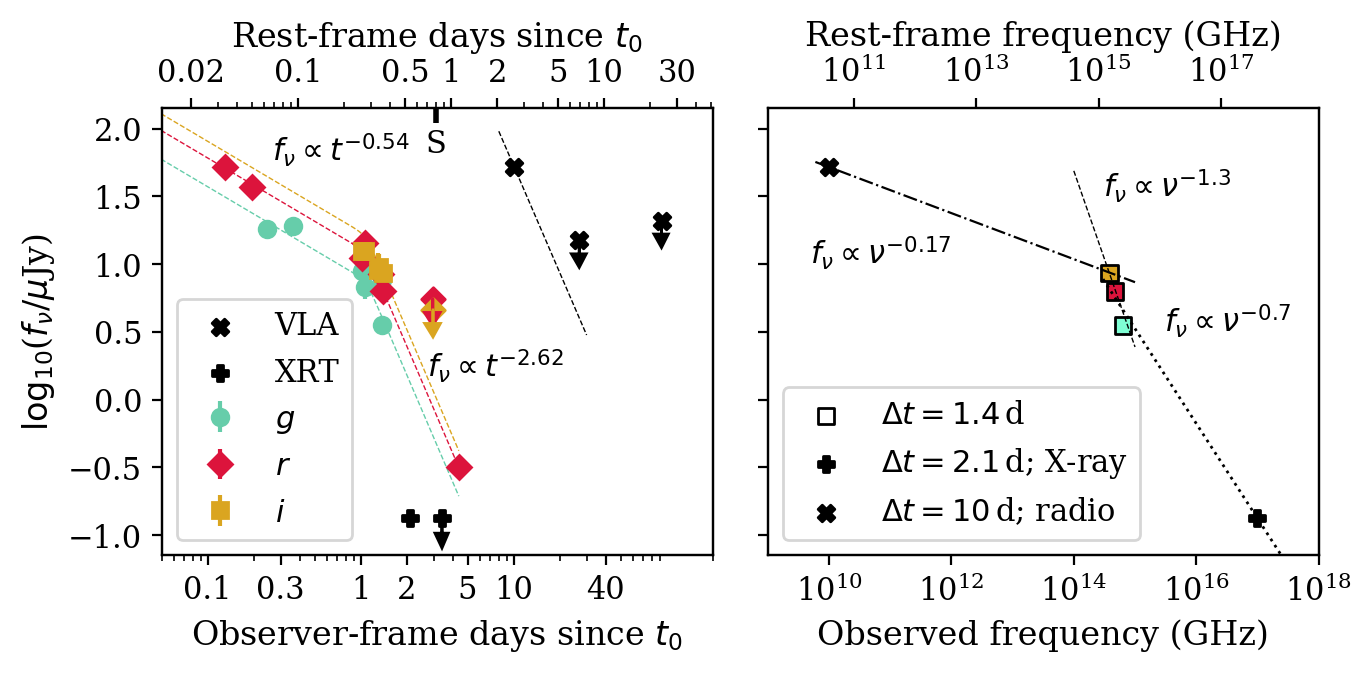}
    \caption{Left: The optical (colored points), X-ray (black plus), and radio (black cross) light curves of AT\,2020blt, shown in observer-frame days on the bottom x-axis and rest-frame days on the top x-axis.
    The X-ray and radio upper limits are at 3-$\sigma$.
    The estimated time of first light $t_0$=Jan 28.18 comes from fitting a broken power-law to the optical light curve (\S\ref{sec:compare-optical-lc}).
    The fitted function is shown as dashed lines.
    For the radio light curve, we show a dotted line with the same temporal index as the post-break optical light curve ($t^{-2.62}$).
    The `S' along the top indicates the epoch of our LRIS spectrum.
    Right: the spectral energy distribution of AT\,2020blt.}
    \label{fig:data}
\end{figure*}

LT $griz$ observations on Jan 29.17
and P60 $gri$ observations 2 hours later
confirmed that AT\,2020blt had red colors.
Furthermore, forced photometry \citep{Yao2019} on P48 images revealed two $g$-band detections that were below the 5-$\sigma$ threshold of the nominal ZTF pipeline, which give $g-r=0.77\pm0.16\,$mag.
Photometry was corrected for Milky Way extinction
following \citet{Schlafly2011} with $E(B-V)=A_V/R_V=0.034\,$mag,
using $R_V=3.1$ and a \citet{Fitzpatrick1999} extinction law.
The full light curve of AT\,2020blt is shown in the left panel of Figure~\ref{fig:data}, and the photometry is listed in Table~\ref{tab:obs}.

\begin{table*}[!ht]
    \centering
    \scriptsize
    \caption{Summary of observations of AT\,2020blt. Time given relative to $t_0$ as defined in \S\ref{sec:compare-optical-lc}. Optical magnitudes have been corrected for Milky Way extinction. P48 values were measured using forced photometry \citep{Yao2019}. X-ray uncertainties are 1-$\sigma$ and upper limits are 3-$\sigma$. Radio upper limits are $3\times$ the image RMS. Uncertainties on radio measurements are given as the quadrature sum of the image RMS and a 5\% uncertainty on the flux density due to flux calibration.}
    \begin{tabular}{ccccc}
    \hline
    \hline
    \multicolumn{5}{c}{Optical Photometry} \\
    \hline
    \textbf{Obs. Date} & \textbf{$\Delta t$ (days)} & \textbf{Instrument} & \textbf{Filter} & \textbf{Mag} \\
Jan 27.54 & -0.64 & P48+ZTF & $r$ & $>21.36$ \\
Jan 28.28 & 0.10 & P48+ZTF & $r$ & $19.60 \pm 0.08$ \\
Jan 28.35 & 0.17 & P48+ZTF & $r$ & $19.97 \pm 0.08$ \\
Jan 28.39 & 0.21 & P48+ZTF & $g$ & $20.74 \pm 0.14$ \\
Jan 28.51 & 0.33 & P48+ZTF & $g$ & $20.70 \pm 0.13$ \\
Jan 29.17 & 0.99 & LT+IO:O & $g$ & $21.52 \pm 0.21$ \\
Jan 29.17 & 0.99 & LT+IO:O & $r$ & $21.29 \pm 0.18$ \\
Jan 29.17 & 0.99 & LT+IO:O & $i$ & $21.15 \pm 0.25$ \\
Jan 29.17 & 0.99 & LT+IO:O & $z$ & $20.82 \pm 0.40$ \\
Jan 29.21 & 1.03 & LT+IO:O & $g$ & $21.83 \pm 0.21$ \\
Jan 29.21 & 1.03 & LT+IO:O & $r$ & $21.00 \pm 0.16$ \\
Jan 29.21 & 1.03 & LT+IO:O & $i$ & $21.15 \pm 0.27$ \\
Jan 29.29 & 1.11 & P60+SEDM & $r$ & $21.03 \pm 0.19$ \\
Jan 29.29 & 1.11 & P60+SEDM & $g$ & $21.30 \pm 0.26$ \\
Jan 29.45 & 1.27 & P48+ZTF & $i$ & $21.52 \pm 0.32$ \\
Jan 29.47 & 1.29 & P48+ZTF & $i$ & $21.45 \pm 0.25$ \\
Jan 29.51 & 1.33 & P48+ZTF & $r$ & $21.58 \pm 0.26$ \\
Jan 29.53 & 1.35 & P200+WaSP & $g$ & $22.53 \pm 0.10$ \\
Jan 29.54 & 1.36 & P48+ZTF & $g$ & $21.69 \pm 0.32$ \\
Jan 29.55 & 1.37 & P200+WaSP & $r$ & $21.90 \pm 0.07$ \\
Jan 29.55 & 1.37 & P200+WaSP & $i$ & $21.56 \pm 0.05$ \\
Jan 31.11 & 2.93 & LT+IO:O & $r$ & $>22.04$ \\
Jan 31.12 & 2.94 & LT+IO:O & $i$ & $>22.24$ \\
Feb 01.53 & 4.35 & Gemini-N+GMOS & $r$ & $25.20\pm0.05$ \\
\hline
\hline
\multicolumn{5}{c}{\textbf{Optical Spectrum with LRIS on Keck-I}} \\
\hline
\textbf{Obs. Date} & \textbf{$\Delta t$ (days)} & \multicolumn{2}{c}{\textbf{Observing Setup}} & \textbf{Exposure Time} \\
Jan 30.64 & 2.44 & \multicolumn{2}{c}{1''-wide slit, 400/3400 grism, 400/8500 grating, D560 dichroic} & 900\,s \\
\hline
\hline
\multicolumn{5}{c}{\textbf{0.3--10\,keV X-ray Observations with \emph{Swift}/XRT}} \\
\hline
\textbf{Obs. Date} & \textbf{$\Delta t$ (days)} & \textbf{Count Rate} & \multicolumn{2}{c}{\textbf{Flux}}  \\
\hline
Jan 29.70 & 2.1 & $(3.96^{+1.30}_{-1.08}) \times 10^{-3} \,\psec$ & \multicolumn{2}{c}{$1.33^{+0.44}_{-0.36} \times 10^{-13}\,\erg\,\psec\,\pcmsq$} \\
Jan 31.04 & 3.4 & $<3.95 \times 10^{-3}\,\psec$ & \multicolumn{2}{c}{$< 1.33 \times 10^{-13}\,\erg\,\psec\,\pcmsq$} \\
\hline
\hline
\multicolumn{5}{c}{\textbf{VLA Radio Observations at 10\,GHz}} \\
\hline
\textbf{Obs. Date} & \textbf{$\Delta t$}  & \textbf{Time On-source}& \textbf{Flux Density} & \textbf{Flux at 10\,GHz} \\
\hline
Feb 07.24 & 10.08 & 0.7\,hr & $52.1 \pm 6.5\,\mu$Jy & $(5.21\pm0.65) \times 10^{-18}\,\erg\,\psec\,\pcmsq$ \\
Feb 23.54 & 26.38 & 0.7\,hr & $<15\,\mu$Jy & 
$<1.5 \times 10^{-18}\,\erg\,\psec\,\pcmsq$ \\
Apr 29.98 & 92.82 & 0.7\,hr & $<21\,\mu$Jy & 
$<2.1 \times 10^{-18}\,\erg\,\psec\,\pcmsq$ \\
\hline
    \end{tabular}
    \label{tab:obs}
\end{table*}

To monitor the light curve,
we triggered a ToO program\footnote{PI: I. Andreoni} with the Wafer-Scale Imager for Prime (WaSP) on the 200-inch Hale telescope at the Palomar Observatory (P200)
and obtained $2\times180\,$s exposures in each of $g$-, $r$-, and $i$-bands.
The WaSP reductions were performed using a pipeline developed for Gattini-IR, described in \citet{De2020}.
The measurement established a rapid fade rate of 2.3 magnitudes in 1.3 days and confirmed the red colors
($g-r=0.63\pm0.12$\,mag).

For a final photometry measurement,
we triggered a ToO observation with the Gemini Multi-Object Spectrograph (GMOS; \citealt{Hook2004}) on the Gemini-North 8-meter telescope on Mauna Kea\footnote{PI: L. Singer; Program ID GN-2019B-Q-130}.
In $8\times200\,$s exposures on Feb 01.53, calibrating against PS1 DR1 \citep{Chambers2016}, we detected the source at $r=25.20\pm0.05\,$mag \citep{Singer2020}.
Data were reduced using DRAGONS (Data Reduction for Astronomy from Gemini Observatory North and South), a Python-based reduction package provided by the Gemini Observatory.
In \S\ref{sec:compare-optical-lc} we model the full optical light curve of AT\,2020blt and compare it to GRB afterglows in the literature.

\subsection{Optical Spectroscopy}
\label{sec:followup-spec}

\begin{figure*}[!hbt]
    \centering
    \includegraphics[width=\linewidth]{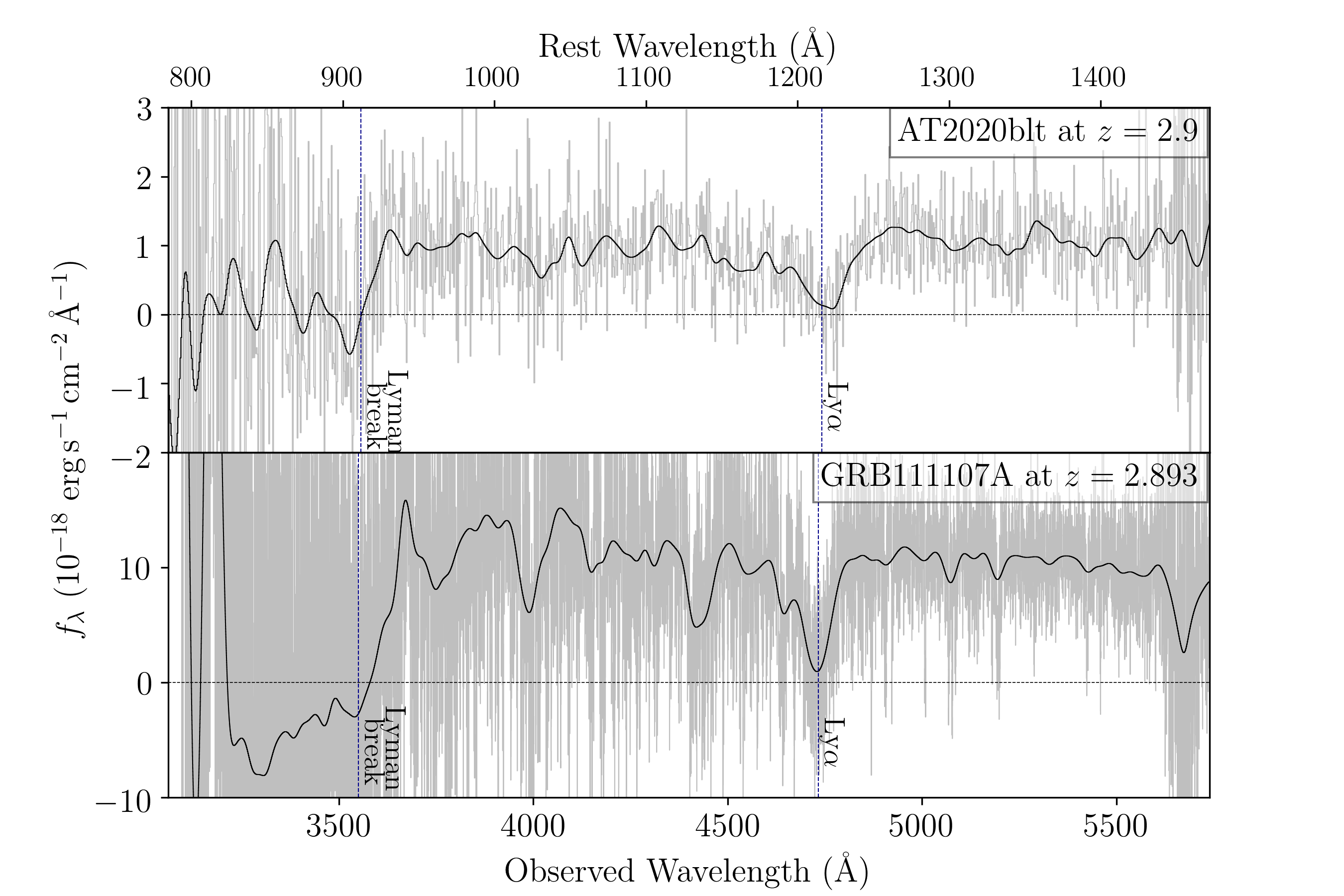}
    \caption{
    Spectrum of AT\,2020blt at $\Delta t=2.4\,\days$ (top panel) with a spectrum of a GRB at a similar redshift in the literature for comparison (bottom panel).
    The spectrum of AT\,2020blt was obtained with the blue arm of LRIS; there was negligible signal in the red arm.
    The spectrum of GRB\,111107A is from \citet{Selsing2019}.
    In each panel, the full spectrum is shown in grey and a smoothed spectrum is overplotted in black. The Lyman-$\alpha$ and Lyman break absorption features are marked with vertical dashed lines.
    We show $f_\lambda=0$ with a horizontal dotted line.}
    \label{fig:spec}
\end{figure*}

We triggered ToO observations\footnote{PI: M. Kasliwal} using the Low Resolution Imaging Spectrometer (LRIS; \citealt{Oke1995}) on the Keck~I 10-m telescope.
The observation details are listed in Table~\ref{tab:obs}.
The spectrum was reduced with \texttt{LPipe} \citep{Perley2019lpipe} and is shown in Figure~\ref{fig:spec}.
The spectrum showed features consistent with the Lyman break (rest-frame 912\,\AA) and Lyman-$\alpha$ absorption (rest-frame 1216\,\AA) at $z = 2.90^{+0.05}_{-0.04}$ (luminosity distance 25\,Gpc\footnote{$\Lambda$CDM cosmology of \citet{Planck2016} used throughout.}),
although the S/N is low due to the short exposure time and the fact that the observation started close to morning twilight.
We searched for narrow lines consistent with
$z = 2.9$ with no convincing detections.
The redshift sets the rest-frame UV magnitude at the time of discovery as $M_{1626\,\text{\AA}}=-25.91\,$mag, assuming a distance modulus of 46.99\,mag and a central wavelength of the ZTF $r$-band filter of 6340\,\AA.

\subsection{X-ray Observations}
\label{sec:followup-xray}

We triggered ToO observations\footnote{PI: A. Ho, Target ID 13197} with the X-Ray Telescope (XRT; \citealt{Burrows2005}) on board
the Neil Gehrels \emph{Swift} observatory \citep{Gehrels2004}. We obtained two epochs of 4\,ks exposures and reduced the data using the online tool\footnote{\url{https://www.swift.ac.uk/user_objects/}} developed by the \emph{Swift} team \citep{Evans2007}.
In the first epoch (Jan 29.70; $\Delta t=2.1\,$d) a source was detected at the position of AT\,2020blt with a 0.3--10\,\kev\ count rate of $(3.96^{+1.30}_{-1.08}) \times 10^{-3} \,\psec$.
Assuming a neutral hydrogen column density $n_H=1.69 \times 10^{20}\,\pcmsq$ and a photon index $\Gamma=2$
the unabsorbed flux density is $1.33^{+0.44}_{-0.36} \times 10^{-13}\,\erg\,\psec\,\pcmsq$.
The source was not detected in the second epoch (Jan 31.04; $\Delta t=3.4\,$d)
with a 3-$\sigma$ confidence upper limit of $<3.95 \times 10^{-3}\,\psec$.
We used \texttt{webpimms}\footnote{\url{https://heasarc.gsfc.nasa.gov/cgi-bin/Tools/w3pimms/w3pimms.pl}} with $\Gamma=2$ and the same value of $n_H$ from the first observation to convert the upper limit on the count rate to an upper limit on the flux density of $< 1.33 \times 10^{-13}\,\erg\,\psec\,\pcmsq$. A log of our X-ray observations is provided in Table~\ref{tab:obs},
and we model the X-ray to radio SED in \S\ref{sec:model-sed}.

\subsection{Radio Observations}
\label{sec:followup-radio}

On Feb 03 we triggered our ToO program on the Karl G. Jansky Very Large Array (VLA; \citealt{Perley2011}) for fast-rising and luminous transients\footnote{VLA/20A-374; PI: A. Ho}.
We obtained an X-band observation on Feb 7.24 (start time; $\Delta t=10.08\,$d)
in C configuration,
using 3C286 as the bandpass and flux density calibrator and J1219+4829 as the phase calibrator.
We calibrated the data using the automated pipeline available in the Common Astronomy Software Applications (CASA; \citealt{McMullin2007}),
and performed additional flagging manually before imaging.
Imaging was performed using the CLEAN algorithm \citep{Hogbom1974} implemented in CASA.
The cell size was 1/10 of the synthesized beamwidth, and the field size was the smallest magic number
($10 \times 2^n$) larger than the number of cells needed to cover the primary beam.
A source was detected at the position of AT\,2020blt with a flux density of $52.1\pm6.5\,\mu$Jy.
In the next X-band image (Feb 23.54; $\Delta t=26.38\,$d)
the source was not detected with an RMS of $5\,\mu$Jy.
In the final observation (Apr 29.98; $\Delta t=92.82\,$d) the source was not detected with an RMS of $7\,\mu$Jy.
A log of our radio observations is provided in Table~\ref{tab:obs}.
In \S\ref{sec:model-sed} we model the X-ray to radio SED and in \S\ref{sec:compare-radio-lc} we put the radio luminosity in the context of GRB afterglows.

\subsection{Search for Associated GRB}
\label{sec:grb-search}

The third Interplanetary Network (IPN\footnote{\url{http://ssl.berkeley.edu/ipn3/index.html}}) consists of six spacecraft that provide all-sky full-time monitoring for high-energy bursts.
The most sensitive detectors in the IPN are the \emph{Swift} Burst Alert Telescope (BAT; \citealt{Barthelmy2005}) the \emph{Fermi} 
Gamma-ray Burst Monitor (GBM; \citealt{Meegan2009}),
and the KONUS instrument on the \emph{Wind} spacecraft \citep{Aptekar1995}.

We searched the \emph{Fermi} GBM Burst Catalog\footnote{\url{https://heasarc.gsfc.nasa.gov/W3Browse/fermi/fermigbrst.html}},
the \fermi-GBM Subthreshold Trigger list\footnote{\url{https://gcn.gsfc.nasa.gov/fermi\_gbm\_subthresh\_archive.html}} (with reliability flag not equal to 2), the
\swift\ GRB Archive\footnote{\url{https://swift.gsfc.nasa.gov/archive/grb\_table/}}, and the Gamma-Ray Coordinates Network archives\footnote{\url{https://gcn.gsfc.nasa.gov/gcn3\_archive.html}} for an associated GRB between the last ZTF non-detection (Jan 27.54) and the first ZTF detection (Jan 28.28).
There were no GRBs coincident with the position and time of AT\,2020blt.\footnote{AT\,2020blt was originally in the localization map of GRB\,200128A because Earth occultation had not been taken into consideration \citep{Hamburg2020}.}

The position of AT\,2020blt was visible\footnote{Search conducted using \url{https://github.com/annayqho/HE_Burst_Search}} to GBM only 65\% of the time: 27\% of the time it was occulted by the Earth, and 8\% of the time GBM was not observing due to a South Atlantic Anomaly passage.
By contrast, KONUS-\emph{Wind} is in interplanetary space, not Earth orbit, and therefore had complete coverage.
KONUS-\emph{Wind} found no detection with a 90\% confidence upper limit on the peak flux of $1.7 \times 10^{-7}\,\erg\,\pcmsq\,\psec$ for a typical long-GRB spectrum\footnote{20--1500\,\kev, 2.944\,s scale, Band spectrum with $\alpha=1, \beta=2.5$, $E_p=300\,\kev$} \citep{Ridnaia2020}.
At the distance of AT\,2020blt, this corresponds to an upper limit on the isotropic gamma-ray luminosity of $L_\mathrm{\gamma,iso} < 1.3 \times 10^{52}\,$erg\,\psec.

Overall, the IPN detects bursts with a 50--300\,keV fluence of 1--3$\times 10^{-6}\,\erg\,\pcmsq$ at 50\% efficiency.
Following \citet{Cenko2013} we take $10^{-6}\,\erg\,\pcmsq$ as a nominal fluence threshold and obtain a limit on the isotropic gamma-ray energy release of $E_\mathrm{\gamma,iso} < 7 \times 10^{52}\,$erg.
We put the limit on $E_\mathrm{\gamma,iso}$ in the context of classical GRBs in \S\ref{sec:analysis}.

\section{Comparison to GRB Afterglows}
\label{sec:analysis}

AT\,2020blt shares a number of features in common with classical GRBs in the literature.
The redshift is typical of GRBs detected
by \emph{Swift} \citep{Gehrels2009} and the absorption features seen in the spectrum are often seen in afterglows at these distances \citep{Fynbo2009,Selsing2019}.
Given the low S/N of our spectrum we are not able to detect common metal lines at this redshift (e.g. \ion{C}{IV}, \ion{Si}{IV}) and we do not attempt to
use the Ly-$\alpha$ absorption feature to measure the host hydrogen column density.

\begin{figure}
    \centering
    \includegraphics[width=\linewidth]{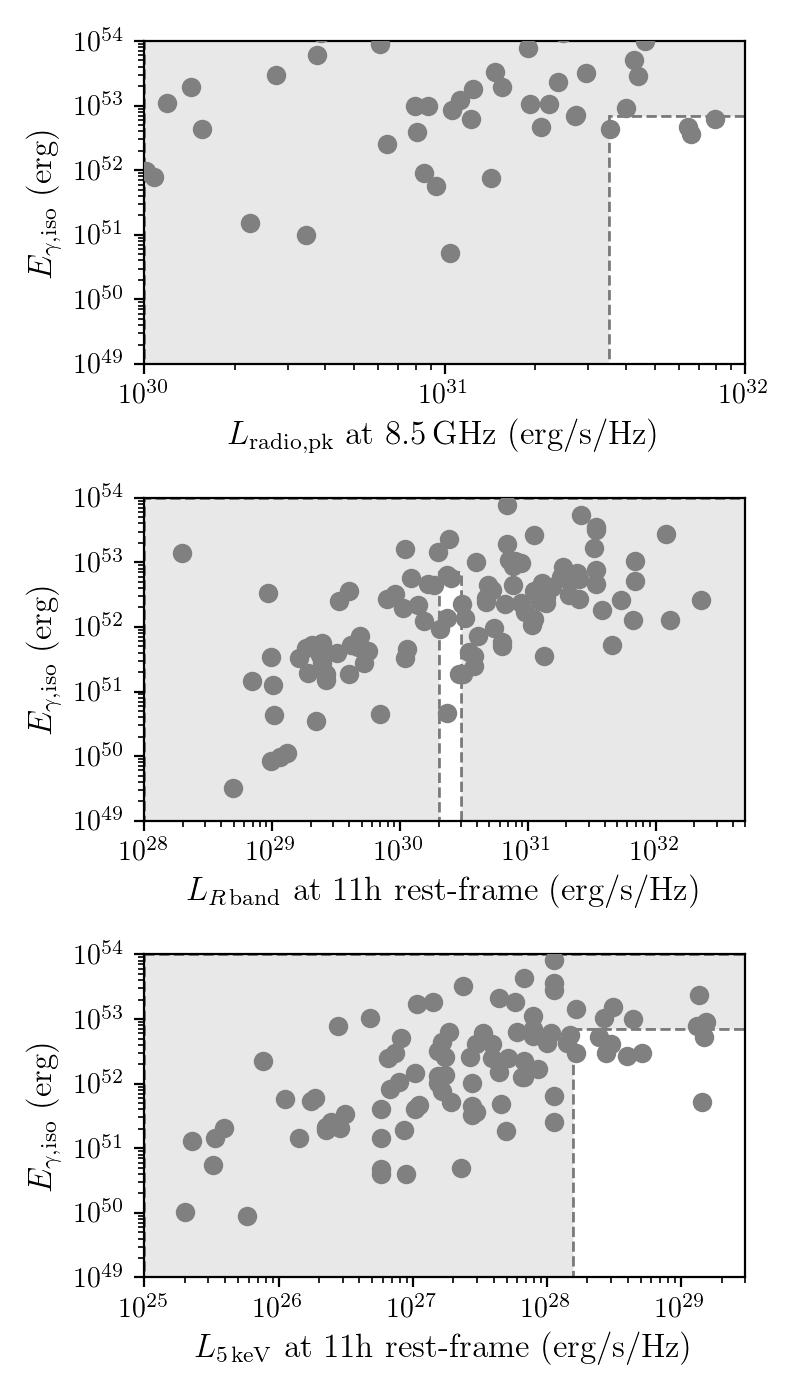}
    \caption{Optical, X-ray, and radio afterglow luminosity of classical GRB afterglows in the literature, compared to the isotropic gamma-ray energy release $E_{\gamma,\mathrm{iso}}$ (grey circles). The optical and X-ray afterglow values were taken from Figure~5 and Figure~6 of \citet{Nysewander2009}, and the radio afterglow values were taken from Figure~20 of \citet{Chandra2012}.
    The region shaded in grey indicates the phase-space ruled out for AT\,2020blt based on our observations and an upper limit on $E_\mathrm{\gamma,iso}$ from KONUS-\emph{Wind}.
    We cannot rule out the possibility that AT\,2020blt was a classical GRB afterglow missed by high-energy detectors.}
    \label{fig:grb-comparison}
\end{figure}

In Figure~\ref{fig:grb-comparison} we compare the X-ray, optical and radio luminosity of AT\,2020blt to GRB afterglows \citep{Nysewander2009,Chandra2012},
and show that a classical GRB cannot be ruled out
based on the limit from KONUS-\emph{Wind} (in general,
at cosmological redshifts KONUS-\emph{Wind} only detects the brightest GRBs).
In the following sections we discuss the optical light curve and SED in more detail.

\subsection{Optical Light Curve}
\label{sec:compare-optical-lc}

As shown in Figure~\ref{fig:data},
the light curve of AT\,2020blt has a clear break well-described by a broken power law.
Optical afterglows with ``classical'' breaks like this were commonly observed prior to the \emph{Swift} era \citep{Kulkarni1999,Harrison2001,Klose2004,Zeh2006},
so it was a surprise when relatively few such breaks were detected in the X-ray afterglows of \emph{Swift} GRBs \citep{Gehrels2009}.
Suggestions for why breaks are rarely detected include that observations do not extend long enough after the burst time \citep{Dai2008}, that the breaks are present in the data but missed in fitting \citep{Curran2008},
that bursts are viewed from a range of viewing angles \citep{Zhang2015},
and that \emph{Swift} GRBs are on average more distant \citep{Gehrels2009} and less energetic \citep{Kocevski2008}.
Furthermore, in X-ray as well as optical bands,
the search for breaks can be complicated by the presence of flares or rebrightening episodes (e.g. \citealt{Kann2010}).

To make a direct comparison to afterglows in the literature with breaks \citep{Zeh2006,Kann2010,Wang2018}
we fit the light curve using a conventional smooth broken power law,
modifying it to take into account the fact that we do not know the burst time $t_0$:

\begin{equation}
\begin{split}
\label{eq:beuermann}
    m(t) &= -2.5 \log_{10}\left( \right. \\ & \left. 10^{-0.4m_c} \left[ 
    \frac{(t-t_0)}{t_b}^{\alpha_1 n} +
    \frac{(t-t_0)}{t_b}^{\alpha_2 n} \right] \right)^{-1/n}.
\end{split}
\end{equation}

In Equation~\ref{eq:beuermann},
$m(t)$ is the apparent magnitude as a function of time,
$n$ parameterizes the smoothness of the break (where $n=\infty$ is a sharp break),
$\alpha_1$ is the power-law index before the break,
$\alpha_2$ is the power-law index after the break,
$t_b$ is the time of the break,
and $m_c$ is the magnitude at the time of the break assuming $n=\infty$.
Note that the original equation also includes terms for the underlying supernova and the host galaxy,
which we take to be zero---a reasonable assumption given that we do not observe any flattening in the optical light curve and the SN would not be detectable in our optical observations at this redshift.

First we fit Equation~\ref{eq:beuermann} to the $r$-band light curve, because it has the most extensive temporal coverage and we cannot necessarily assume constant colors across the optical light curve.
Using the Levenberg-Marquardt algorithm implemented in \texttt{scipy} we find parameters that are very poorly constrained:
$m_c = 20.99 \pm 5.01$\,mag, $t_0 = 2458876.69\pm0.42$, $t_b=1.00\pm1.84\,$d, $\alpha_1 = 0.52 \pm 2.81$, $\alpha_2 = 2.59 \pm 0.26$.
The fit has a reduced $\chi^2/\nu=3.6/\nu$ where $\nu=1$ is the number of degrees of freedom (number of data points minus number of fitted parameters).

In \S\ref{sec:disc-basic-analysis} we show that constant colors are a reasonable assumption at optical frequencies.
So, to obtain more precise parameters we fit
Equation~\ref{eq:beuermann} to the
$g$-, $r$-, and $i$-band light curves simultaneously, assuming constant $g-r$ and $r-i$ offsets.
The result is
$m_{c,r} = 20.96 \pm 0.53$\,mag, 
$m_{c,g} = 21.50 \pm 0.53$\,mag, 
$m_{c,i} = 20.66 \pm 0.53$\,mag, 
$t_0 = 2458876.68\pm0.01$, $t_b=1.00\pm0.10\,$d, $\alpha_1 = 0.54 \pm 0.26$, and $\alpha_2 = 2.62 \pm 0.01$. The smoothing parameter is still poorly constrained: $n=4\pm67$.
Note that we do not assume a single spectral index across the optical band because the $g$-band flux is slightly attenuated by the Ly-$\alpha$ absorption feature and the Lyman forest (from the spectrum we estimate that it is attenuated at the 15\% level, corresponding to about 0.2\,mag).
The fit has a reduced $\chi^2/\nu=11.8/\nu=1.31$ where $\nu=10$ is the number of degrees of freedom.
Throughout the paper, we use the parameters resulting from this multi-band fit, which results in a best-fit light curve shown in the left panel of Figure~\ref{fig:data}.

\begin{figure}
    \centering
    \includegraphics[width=\linewidth]{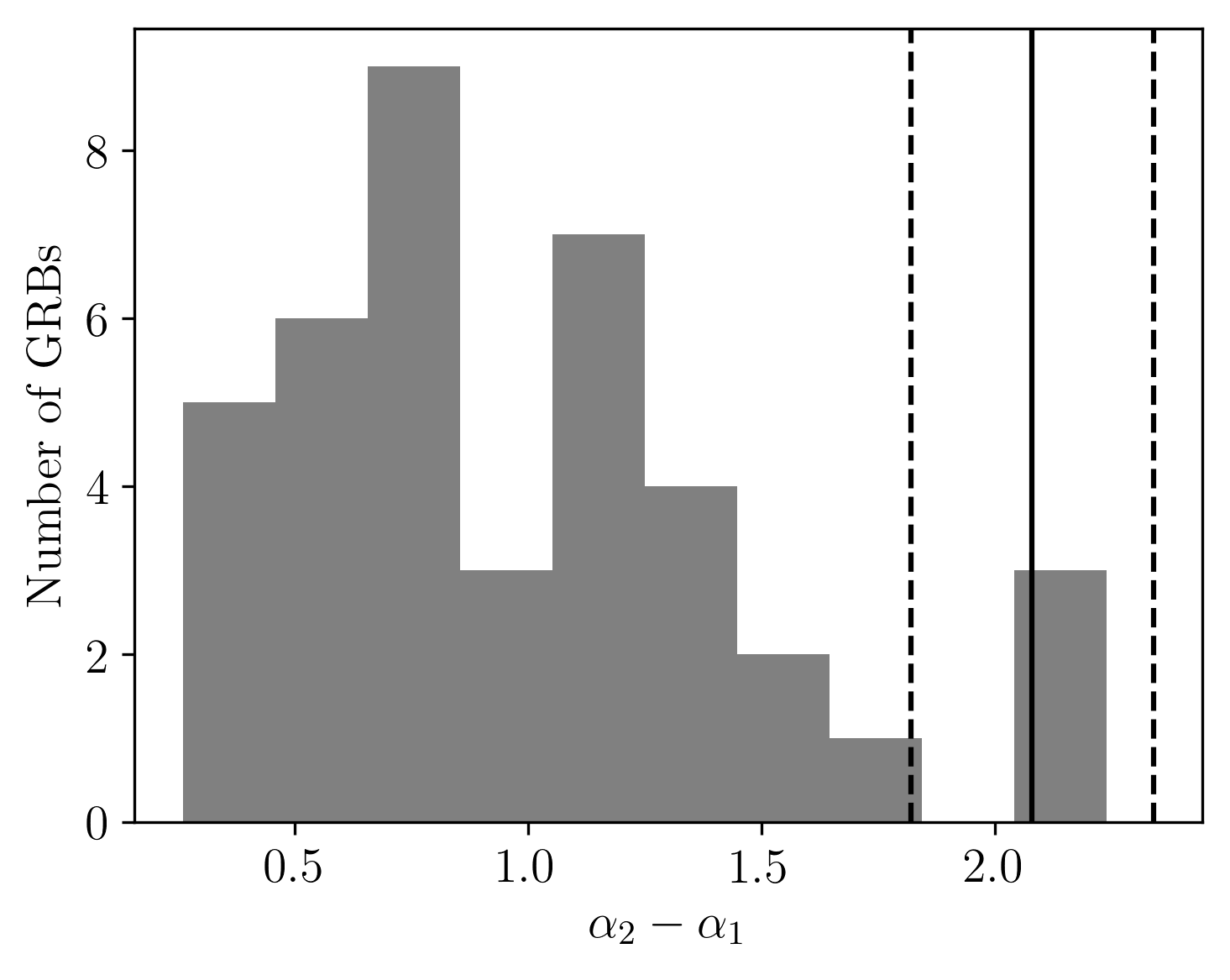}
    \caption{The difference between the post-break and pre-break temporal index, compared to a sample of GRBs with jet breaks from the literature \citep{Zeh2006,Kann2010}.
    The solid vertical line is the best-fit value of $\alpha_2-\alpha_1$ from \S\ref{sec:compare-optical-lc}.
    The dashed vertical lines represent the error bars on the best-fit value.
    }
    \label{fig:alphahist}
\end{figure}

The best-fit $t_0$ is Jan 28.18 $\pm$ 0.01,
which is 2.4 hours before the first detection and 15.6 hours after the last non-detection.
The best-fit $t_\mathrm{j}=1.00\pm0.10\,$d after $t_0$ (observer-frame)
is typical of optical afterglows with breaks (e.g. \citealt{Zeh2006,Kann2010,Wang2018}).
In Figure~\ref{fig:alphahist} we show the resulting value of $\Delta \alpha = 2.08\pm0.26$ compared to the distribution in \citet{Zeh2006} and \citet{Kann2010}.
The value of $\Delta \alpha$ appears to be large compared to afterglows in the literature,
but not unprecedented.

The origin of breaks in afterglow light curves is still debated.
A leading hypothesis is that a break results from a collimated jet \citep{Rhoads1997,Sari1999}.
The traditional argument is that while $\Gamma(t) \gg \theta^{-1}$, the emission cannot be distinguished from an isotropic outflow, because relativistic beaming confines the viewing angle to a small region that is expanding too quickly to interact sideways.
As $\Gamma(t)$ decreases to $\Gamma(t) \sim \theta^{-1}$, two effects become important:
the jet begins expanding sideways \citep{Rhoads1997},
and the edge of the jet becomes visible \citep{Meszaros1999}.
However, ``textbook'' achromatic breaks are rarely observed,
and simulations suggest that breaks can be chromatic \citep{vanEerten2011} and that sideways expansion can take place significantly later than when the edge of the jet becomes visible \citep{Panaitescu1998,Granot2012}.

If the break in the light curve of AT\,2020blt is a jet break --- and we caution that it is rare to see breaks that actually behave in the way one would expect for jet breaks, e.g. \citet{Liang2008} --- we can use the timing of the break to estimate the opening angle of the jet $\theta_0$. For a constant-density ISM we have from \citet{Sari1999} that

\begin{equation}
    t_\mathrm{jet} \approx 6.2 (E_{52}/n_1)^{1/3}(\theta_0/0.1)^{8/3}\,\mathrm{hr}
\end{equation}

\noindent where $E_{52}$ is the kinetic energy release of the explosion in units of $10^{52}\,$erg, $\theta_0$ is in radians, and $n_1$ is the ambient density in units $1\,\pcmcub$.
Using our rest-frame value $t_\mathrm{jet}=0.26\pm0.03\,$d ($6.24\pm0.62$\,hr)
we have

\begin{equation}
    1.0 \pm 0.1 = (E_{52}/n_1)^{1/3}(\theta_0/0.1)^{8/3}\,\mathrm{hr}.
\end{equation}

\noindent 
From the GRB search in \S\ref{sec:grb-search} we have an approximate upper limit on the isotropic gamma-ray energy release of $E_{52} < 7$.
The $E_{52}/n_1$ term has a much weaker dependence than the opening angle term,
and can be reasonably estimated to be unity
based on typical GRB environment densities \citep{Chandra2012}.
We find an opening angle of $\theta_0 = 0.10 \pm 0.04 = 5.7\pm2.3$\,degrees,
typical of opening angles inferred from optical jet breaks \citep{Panaitescu2001,Zeh2006,Wang2018}.

If the break is due to the material spreading sideways,
then the temporal index after the break is
$F_\nu(t) \propto t^{-p}$ \citep{Sari1999}
where $p$ is the power-law index of the electron energy distribution.
Using the value of $\alpha_2$ above we have
$p=\alpha_2=2.62\pm0.01$,
which is large but consistent with values expected for shock acceleration \citep{Jones1991} given the uncertainties, and furthermore is within the normal range of values inferred from optical afterglows in the literature \citep{Wang2018}.

If the steepening is due solely to detecting the edge of the jet, the expected post-break slope\footnote{Here we assume that the optical frequency $\nu$ is in the regime $\nu_m<\nu<\nu_c$, motivated in \S\ref{sec:model-sed} and \S\ref{sec:compare-radio-lc}.} is the slope of a spherically expanding outflow ($t^{-3(p-1)/4}$; \citealt{Sari1999}) with two additional powers of $\Gamma \propto t^{-3/8}$.
The resulting temporal slope is $F_\nu(t) \propto t^{-3p/4}$,
so $3p/4=\alpha_2=2.62\pm0.01$.
The value of $p=3.49\pm0.01$ is larger than what is predicted for shock acceleration \citep{Jones1991} so in what follows we assume $p=2.62\pm0.01$.

\subsection{Spectral Energy Distribution}
\label{sec:model-sed}

The spectrum of an afterglow is determined by the kinetic energy of the explosion, the ambient density,
and the fraction of the energy in magnetic fields $\epsilon_B$ and relativistic electrons $\epsilon_e$ \citep{Sari1998}.
The spectrum is characterized by several break frequencies:
the cooling frequency $\nu_c$, the characteristic frequency $\nu_m$, and the self-absorption frequency $\nu_a$.
The spectral index in any region of the spectrum therefore depends on physical properties of the explosion and on the position of the observing frequency relative to the break frequencies.

In \S\ref{sec:compare-optical-lc} we found $p=2.62\pm0.01$ based on the post-break light curve power-law index.
Because the $g$-band measurement is attenuated by Lyman-$\alpha$ and Lyman forest absorption,
we use the $r-i$ color from the WASP observation
($r-i=0.34\pm0.09\,$mag) to estimate that the optical spectral index $\beta_\mathrm{opt} = 1.3\pm0.4$.
Such a steep spectral index is only ever observed as a result of absorption \citep{Cenko2009,Greiner2011}:
at this redshift even $i$-band is well into the far-ultraviolet, so it takes relatively little extinction to significantly alter the flux and color. 
If $\beta_\mathrm{opt}$ were the ``true'' (unextincted) spectral index, that would indicate that the cooling frequency $\nu_c$ lies below the optical bands \citep{Sari1999}. Taking $\nu_c < 10^{14}\,$Hz at $t_d \approx 0.5\,$d we have, following \citet{Sari1998}

\begin{equation}
    26 > \epsilon_B^{-3/2} E_{52}^{-1/2} n_1^{-1}.
\end{equation}

\noindent Using $\epsilon_B < 3 \times 10^{-4}$ and $E_{52}=1$ (\S\ref{sec:compare-radio-lc}) we find a very large CSM density of $n > 7 \times 10^{3}\,\pcmcub$,
which would violate the assumption that we made in estimating the jet opening angle of $E_{52}/n_1\approx1$.
Furthermore, as shown in the right panel of Figure~\ref{fig:data}, the optical to X-ray spectral index is $\beta_{\mathrm{opt},X}\approx 0.7$, inconsistent with $\beta_{\mathrm{opt}}$.
It seems natural that the optical spectral index is steepened by dust attenuation,
and $\beta_{\mathrm{opt},X}$ is the ``true'' spectral index, rather than $\beta_{\mathrm{opt}}$.

The value of $\beta_{\mathrm{opt},X}$ is consistent with $(p-1)/2$ but not with $p/2$, so the cooling frequency $\nu_c$ lies above the X-ray band \citep{Sari1999}. For adiabatic evolution we have \citep{Sari1998}

\begin{equation}
    \nu_c = 2.7 \times 10^{12} \epsilon_B^{-3/2} E_{52}^{-1/2} n_1^{-1} t_d^{-1/2}\,\hz.
\end{equation}

\noindent where $t_d$ is the time (in days) after the explosion. Taking $\nu_c > 10^{18}\,\hz$ and $t_d=0.5\,\days$ (time of the X-ray observation; rest-frame), we have

\begin{equation}
    2.6 \times 10^{5} < \epsilon_B^{-3/2} E_{52}^{-1/2} n_1^{-1}.
\end{equation}

\noindent Using $\epsilon_B < 3 \times 10^{-4}$ and $E_{52}=1$ (\S\ref{sec:compare-radio-lc}) we find $n < 0.7\,\pcmcub$, which is typical for GRBs \citep{Panaitescu2001,Chandra2012}.

\subsection{Radio Light Curve}
\label{sec:compare-radio-lc}

In \S\ref{sec:compare-optical-lc} we showed that the optical light curve of AT\,2020blt is fairly typical for classical GRBs.
However, the radio light curve of AT\,2020blt (left panel of Figure~\ref{fig:data}) decays steeper than
$F_\nu \propto t^{-2.3}$, which is unusual for
GRBs with detected radio afterglows in general \citep{Chandra2012},
including PTF11agg \citep{Cenko2013}.

Early fast-evolving emission in GRB radio afterglow light curves can arise from reverse shocks or diffractive scintillation in the interstellar medium \citep{Laskar2013,Perley2014,Laskar2016,Alexander2017,Laskar2018,Alexander2019}.
To determine whether scintillation could be the origin in this case,
we use the NE2001 model of the ISM \citep{Cordes2002}.
For context, scintillation results from small-scale inhomogeneities in the ISM, which change the phase of an incoming wavefront.
As the Earth moves, the line of sight to a background source changes,
so the net effect is an observed change in flux.
The effect is greatest for sources observed at a frequency $\nu_\mathrm{obs}$ that is close to the transition frequency $\nu_0$, which separates strong scattering ($\nu_\mathrm{obs}<\nu_0$) from weak scattering ($\nu_\mathrm{obs}>\nu_0$).

Using the NE2001 map,
we determine that the line-of-sight towards AT\,2020blt has a transition frequency $\nu_0=7.12\,$GHz,
which is close to our observing frequency.
Furthermore, we can estimate the timescale for flux changes.
Using $D=100$\,pc as the characteristic scale height of the ISM, and $\lambda=3$\,cm as our observing wavelength,
the Fresnel length is $r_f = \sqrt{\lambda D} \approx 10^{10}\,$cm.
Assuming that Earth moves at $v=30\km\,\psec$, we obtain $t \sim r_f/v\approx1\,$hr.
In conclusion, the flux could easily change by an order of magnitude due to scintillation over the large time window (16\,d, observer-frame) between our observations.
Note that the timescale is close to our
time on-source ($\sim30$\,minutes),
so there could be some damping of the scintillation over the course of our observation.
However, the signal-to-noise of the data is not high enough for us to search for scintillation within the observation.

If, on the other hand, the rapid change in flux is due to a truly steep power-law decay in the radio emission,
there would be implications for the ambient density and the value of $\epsilon_B$.
In particular, the characteristic frequency $\nu_m$ must lie below the radio band. For adiabatic evolution we have \citep{Sari1998}

\begin{equation}
    \nu_m = 5.7 \times 10^{14} \epsilon_B^{1/2} \epsilon_e^2 E_{52}^{1/2} t_d^{-3/2}\,\hz.
 \end{equation}
 
Requiring $\nu_m < 10\,\ghz$ (39\,\ghz\ rest-frame) at $t_d=2\,\days$ (time of the radio detection; rest-frame) and adopting $\epsilon_e=0.1$ \citep{Kumar2015,Beniamini2017} we find 

\begin{equation}
    0.02 > \epsilon_B^{1/2} E_{52}^{1/2}.
 \end{equation}
 
Assuming $E_{52}=1$, we find $\epsilon_B < 3 \times 10^{-4}$, which is also typical for GRBs based on high-energy and optical afterglow modeling \citep{Kumar2015,Beniamini2017}.
So, although an early steep-decaying radio light curve is unusual for GRBs with detailed radio observations, we have no reason to believe that the radio behavior of AT\,2020blt is unusual for the population of GRBs as a whole.

\section{Interpretation}
\label{sec:interpretation}

In \S\ref{sec:analysis} we found that the optical and radio light curve of AT\,2020blt is similar to that of classical GRB afterglows.
The fact that we observed an achromatic steepening suggests that there was a jet break,
which requires that our observing angle was within the jet opening angle.

Three possibilities remain for the origin of AT\,2020blt.
The first (and simplest) possibility is that AT\,2020blt was a classical GRB viewed directly on-axis ($\theta_\mathrm{obs} < \theta_0$) for which the high-energy emission was simply missed by GRB satellites.
As discussed in \S\ref{sec:analysis},
the on-axis scenario is entirely possible.
With an eye to the future,
when larger samples of optical afterglows will be available (including some with more stringent limits on associated GRB emission),
we consider two additional possibilities:
that AT\,2020blt is a classical GRB observed slightly off-axis $\theta_\mathrm{obs} \gtrsim \theta_0$ (\S\ref{sec:interp-slightly-off-axis}) and that AT\,2020blt is a dirty fireball (\S\ref{sec:dirty-fireball}).

\subsection{A Slightly Off-Axis GRB}
\label{sec:interp-slightly-off-axis}

Here we consider the possibility that AT\,2020blt was a classical GRB viewed slightly outside the jet opening angle.
\citet{Beniamini2019} argued that the vast majority of GRBs observed so far must have been observed close to or within the jet core, implying that GRB emission is not produced efficiently away from the core.
So, as discussed in \S\ref{sec:introduction}, there is a natural expectation for X-ray and optical afterglows without detected GRB emission \citep{Meszaros1997,Rhoads1997,Nakar2003}.
The slightly off-axis model has been invoked to explain low-luminosity GRBs or X-ray flashes \citep{RamirezRuiz2005} as well as plateaus observed in X-ray afterglow light curves \citep{Eichler2006,Beniamini2020_xray}.

One signature of a slightly off-axis afterglow could be an early shallow decay and a large value of $\Delta \alpha$ \citep{Ryan2019,Beniamini2020}.
This can be understood as follows.
In on-axis events, the early stage of the light curve is set by two competing effects: the shock is decelerating, but the beaming cone is widening to include more material.
In a slightly off-axis event, there is a third effect, which is that the beaming cone widens to include material of increasing energy per solid angle---hence a shallower decay.

As shown in Fig.~\ref{fig:alphahist}, we did observe a large value of $\Delta \alpha$ in AT\,2020blt.
A larger number of events would help to test this hypothesis:
the luminosity function of the early afterglow should be different from the luminosity function of directly on-axis afterglows,
and the distribution of limits on $E_\mathrm{iso}$ would eventually make it unlikely that the afterglows are drawn from the same population as classical GRBs.
With more events, we could hope to make the first measurement of the optical beaming factor in GRB afterglows \citep{Nakar2003}.

\subsection{A Dirty Fireball}
\label{sec:dirty-fireball}

Here we consider the possibility that AT\,2020blt was a ``dirty fireball,'' i.e. a jet with lower Lorentz factor ($\Gamma \sim 10$) that did not produce any GRB emission, as proposed for PTF11agg \citep{Cenko2013}.
The basis for the dirty fireball argument for PTF11agg was the rate: at the time, it seemed that the rate of PTF11agg-like events may have been significantly higher than the rate of classical GRBs \citep{Cenko2013}, although this was later shown to not be the case \citep{Cenko2015,Ho2018}.
Taking a similar approach to \citet{Cenko2015} and \citet{Ho2018},
we searched high-cadence (6$\times$/night) ZTF survey data \citep{Bellm2019surveys} from 2018 March 1 to 2020 May 12 to estimate the areal exposure in which an event like AT\,2020blt would have passed our filter.

We folded the light curve of AT\,2020blt through all $r$-band exposures in the ZTF high-cadence fields, varying the burst time by 0.01\,d intervals, to see over what duration the transient would have had two $r$-band detections above the limiting magnitude,
with a first detection over one magnitude brighter than the last non-detection.
We found a total exposure of 855 field-nights.
We assume a 100\% detection efficiency, so our result is somewhat of a lower limit, particularly at these fainter magnitudes; the efficiency as a function of limiting magnitude has not yet been characterized for ZTF.
The 92 high-cadence survey fields included in our search have a combined footprint of 3307\,\degsq\ after removing the overlap between fields.
So, we estimate the all-sky rate of transients similar to AT\,2020blt to be

\begin{equation}
\begin{split}
    {\cal R} & \equiv \frac{N_\mathrm{rel}}{A_\mathrm{eff}} \\
    & = \frac{1}{30,734\,\degsq\,\days} \times \frac{365.25\,\days}{\mathrm{year}} 
    \times \frac{41,253\,\degsq}{\mathrm{sky}} \\
    & = 490\,\pyr
\end{split}
\end{equation}

\noindent with a 68\% confidence interval from Poisson statistics of 85--1611\,\pyr.
For comparison, the all-sky rate of \emph{Swift} GRBs out to $z=3$ has been estimated to be $1455^{+80}_{-112}\,\pyr$ \citep{Lien2014}.
The \emph{Swift} GRB rate is larger than the rate of optical afterglows,
since only a subset of GRBs show bright optical afterglow emission \citep{Cenko2009}.
So, within the uncertainties, the rate of optical afterglows in ZTF is compatible with the GRB rate.
We therefore concur with the conclusion in
\citet{Cenko2015} and \citet{Ho2018} that there is no evidence for an afterglow-like phenomenon that is significantly more common than classical GRBs.

The light curve of a dirty fireball will take longer to rise to peak because it is set by the time it takes the shock to sweep up material of mass $1/\Gamma_0$ times the ejecta mass (the ``deceleration'' time).
For a uniform-density medium, the expression (observer-frame) is

\begin{equation}
t_\mathrm{dec} = 30\,E_{53}^{1/3} n^{-1/3} \Gamma_{0,2.5}^{-8/3}\,\sec.
\end{equation}

So, an outflow with $\Gamma_0=100$ will have an afterglow that rises to peak in 300\,s, but an outflow with $\Gamma_0=10$ will have an afterglow that rises to peak in 1.2\,d.
The power-law index of the rising light curve will remain the same, however.
We searched for GRBs within three days prior to the discovery of AT\,2020blt, but found no coincident bursts.
So, we have no evidence for such a long rise time in AT\,2020blt.

\section{Summary and Conclusions}
\label{sec:conclusions}

To summarize, we used a filter for extragalactic fast transients
together with fast-turnaround follow-up observations to discover a cosmological afterglow ($z \approx 2.9$) in ZTF high-cadence data.
Our search strategy (\S\ref{sec:disc-ztfdiscovery}) is to find fast-appearing transients with no host galaxy and red colors, inconsistent with the thermal emission expected for the foreground fog of stellar flares. Additional photometry obtained within 24 hours confirmed rapid fading, and a spectrum obtained within three days established the cosmological origin (\S\ref{sec:followup}).
AT\,2020blt is one of only a few optical afterglows discovered independently of a high-energy trigger,
and one of only two events with both a redshift measurement and no detected GRB.

One lesson from our work is that for a single event, it is very difficult to rule out a classical GRB missed by high-energy detectors.
The most sensitive detectors have the smallest probability of observing the field over the relevant time interval, given the typical cadence of optical observations.

We consider what might be possible with a large sample of events.
From existing survey data, it is already clear that the rate of afterglow-like events cannot be significantly higher than the rates of classical GRBs (\S\ref{sec:interpretation}).
Dirty fireballs could have a significantly longer duration (\S\ref{sec:dirty-fireball}),
in which case they would not pass our fast-transient filter and the rate could be significantly higher than the limits set by intra-night fast-transient searches.
The appearance of slightly off-axis events (\S\ref{sec:interp-slightly-off-axis})
depends on the structure of the jet, currently unknown,
but the luminosity function should be different (with lower overall luminosity) than that of classical GRBs.

Perhaps the strategy of searching for intra-night transients is too restrictive.
A more agnostic strategy could be to
search for relativistic explosions on the basis of luminosity.
If dirty fireballs have an intrinsically lower redshift distribution,
then their host galaxies are more likely to be detected;
in fact, of the three afterglows with ZTF detections,
two (ZTF19aabgebm and ZTF19abvizsw) have detected host galaxies in the Legacy Survey DR8 \citep{Dey2019} with high photometric redshifts.
In a search for luminous transients, interlopers like superluminous supernovae could be easily ruled out by light-curve duration.
A search for luminous transients using host-galaxy photometric and spectroscopic catalogs during ZTF Phase II could help set the stage for a similar search strategy during LSST.

\vspace{5mm}
\facilities{Swift, EVLA, VLA, Liverpool:2m, PO:1.2m, PO:1.5m}

\software{{\tt CASA} \citep{McMullin2007},
          {\tt astropy} \citep{Astropy2013,Astropy2018},
          {\tt matplotlib} \citep{Hunter2007},
          {\tt scipy} \citep{Virtanen2020},
          {\tt DRAGONS}
}

\acknowledgements

It is a pleasure to thank the anonymous referee for a thorough and thoughtful report that greatly improved the quality of the paper.
A.Y.Q.H. would like to thank
Udi Nakar for pointing out that dirty fireballs will have a longer rise time than clean fireballs,
and Chris Bochenek and Vikram Ravi for useful discussions regarding scintillation of radio point sources.
She would also like to thank Steve Schulze, Eran Ofek, and Avishay Gal-Yam, and David Kaplan for their detailed reading of the manuscript.

A.Y.Q.H. and K.D. were supported
by the GROWTH project funded by the National Science Foundation under PIRE Grant No. 1545949.
A.Y.Q.H. was also supported by the Miller Institute for Basic Research in Science at the University of California Berkeley.
A.~A.~Miller is funded by the Large Synoptic Survey Telescope Corporation, the
Brinson Foundation, and the Moore Foundation in support of the LSSTC Data
Science Fellowship Program; he also receives support as a CIERA Fellow by the
CIERA Postdoctoral Fellowship Program (Center for Interdisciplinary
Exploration and Research in Astrophysics, Northwestern University).
C.F.~gratefully acknowledges support of his research by the Heising-Simons Foundation (\#2018-0907).
A.~Goobar acknowledges support from the K \& A Wallenberg Foundation, the Swedish Research Council (VR),  and the GREAT research environment grant 2016-06012.

Based on observations obtained with the Samuel Oschin Telescope 48-inch and the 60-inch Telescope at the Palomar Observatory as part of the Zwicky Transient Facility project. ZTF is supported by the National Science Foundation under Grant No. AST-1440341 and a collaboration including Caltech, IPAC, the Weizmann Institute for Science, the Oskar Klein Center at Stockholm University, the University of Maryland, the University of Washington, Deutsches Elektronen-Synchrotron and Humboldt University, Los Alamos National Laboratories, the TANGO Consortium of Taiwan, the University of Wisconsin at Milwaukee, and Lawrence Berkeley National Laboratories. Operations are conducted by COO, IPAC, and UW.
SED Machine is based upon work supported by the National Science Foundation under Grant No. 1106171.
This work made use of data supplied by the UK Swift Science Data Centre at the University of Leicester.
Based on observations obtained at the international Gemini Observatory, a program of NSF’s OIR Lab, which is managed by the Association of Universities for Research in Astronomy (AURA) under a cooperative agreement with the National Science Foundation. on behalf of the Gemini Observatory partnership: the National Science Foundation (United States), National Research Council (Canada), Agencia Nacional de Investigaci\'{o}n y Desarrollo (Chile), Ministerio de Ciencia, Tecnolog\'{i}a e Innovaci\'{o}n (Argentina), Minist\'{e}rio da Ci\^{e}ncia, Tecnologia, Inova\c{c}\~{o}es e Comunica\c{c}\~{o}es (Brazil), and Korea Astronomy and Space Science Institute (Republic of Korea).
Gemini data were processed using the Gemini IRAF package and DRAGONS (Data Reduction for Astronomy from Gemini Observatory North and South).
The Liverpool Telescope is operated on the island of La Palma by Liverpool John Moores University in the Spanish Observatorio del Roque de los Muchachos of the Instituto de Astrofisica de Canarias with financial support from the UK Science and Technology Facilities Council.

\bibliography{refs}
\bibliographystyle{aasjournal}

\end{document}